\documentclass[]{article}   
\usepackage{fullpage}

\usepackage[]{graphicx}
\usepackage{amsmath}
\usepackage{amssymb}
\usepackage{epsfig}
\usepackage{color}
\usepackage{subfigure}

\def\##1{{\bf #1}}
\def\=#1{\underline{\underline #1}}
\def\~#1{{\tilde{\bf #1}}}

\def\r#1{(\ref{#1})}

\def\epso{\epsilon_{\scriptscriptstyle 0}}

\def\muo{\mu_{\scriptscriptstyle 0}}
\def\ko{k_{\scriptscriptstyle 0}}
\def\etao{\eta_{\scriptscriptstyle 0}}
\def\co{c_{\scriptscriptstyle 0}}

\def\inc{_{\rm inc}}
\def\sca{_{\rm sca}}
\def\exc{_{\rm exc}}
\def\ext{_{\rm ext}}
\def\abs{_{\rm abs}}

\def\eqvt{_{\rm eqvt}}

\def\smn{_{smn}}

\def\les{\left[}
\def\ris{\right]}
\def\lec{\left\{}
\def\ric{\right\}}

\def\.{\mbox{ \tiny{$^\bullet$} }}

\def\ux{\hat{\#x}}
\def\uy{\hat{\#y}}
\def\uz{\hat{\#z}}

\def\ur{\hat{\#r}}

\def\st{\sin\theta}
\def\ct{\cos\theta}
\def\sp{\sin\phi}
\def\cp{\cos\phi}

\begin{document}

\begin{center}
{\bf Light scattering by a vacuum-like sphere with   magnetoelectric gyrotropy}\\

{A. D. Ulfat Jafri${}^{a}$  and  Akhlesh Lakhtakia${}^{b,\ast}$}\\

{$^a$Department of Electronics,
Quaid-i-Azam University, Islamabad, Pakistan}\\
{$^b$NanoMM---Nanoengineered Metamaterials Group,
Department of Engineering Science and Mechanics, Pennsylvania State University,\\
University Park, PA 16802, USA}\\
{$^\ast$Corresponding author: akhlesh@psu.edu}
\end{center}

\begin{abstract}
An exact transition matrix was formulated for electromagnetic scattering by a vacuum-like sphere with  magnetoelectric gyrotropy.
Both the total scattering and forward scattering efficiencies are lower when the magnetoelectric gyrotropy vector of the sphere is co/anti-parallel to the electric field or magnetic field of an incident plane wave than when the magnetoelectric gyrotropy vector is coparallel to the propagation vector of the incident plane wave.
Backscattering is absent when the propagation vector is co/anti-paralel to the magnetoelectric gyrotropy vector.
\end{abstract}

 \section{Introduction}

The introduction of research on metamaterials has fuelled hope that  the propagation of light
in certain gravitational scenarios can be emulated in the laboratory \cite{CK2009,ML2010,CML2010,Smoly2011,Smoly2013},
possibly as  composite materials \cite{Lu,MLprb2011}.
The theoretical basis for this hope lies in the identification \cite{Plebanski,MLSnjp2004} of the components of the gravitational metric $g_{\alpha\beta}$, $\alpha\in\lec0,1,2,3\ric$
and $\beta\in\lec0,1,2,3\ric$,  with the components of the constitutive
dyadics $\=\gamma$ and ${\bf\Gamma}\times\=I$ of a linear bianisotropic continuum specified as
\begin{equation}
\label{conrel0}
\left.\begin{array}{l}
{\bf D} =\epso\=\gamma\.{\bf E} -\co^{-1}({\bf\Gamma}\times\=I)\.{\bf H}
\\
{\bf B} =\muo\=\gamma\.{\bf H} +\co^{-1}({\bf\Gamma}\times\=I)\.{\bf E}
\end{array}\right\}\,,
\end{equation}
where $\=I$ is the identity dyadic;
$\epso$ and $\muo$ are the permittivity and the permeability of free space, respectively; and
$\co=1/\sqrt{\epso\muo}$.  Specifically,
when the metric  $g_{\alpha\beta}$ has $(+,-,-,-)$ as its  signature and $\bar{g}$ denotes the determinant of  $g_{\alpha\beta}$, we get $\gamma_{\ell{m}}=-({-\bar{g}})^{1/2}g^{\ell{m}}/g_{00}$
and $\Gamma_\ell=g_{0\ell}/g_{00}$,  $\ell\in\lec1,2,3\ric$
and $m\in\lec1,2,3\ric$. Thus, bianisotropic electromagnetics \cite{CK1968,MLbook} is already a theoretical testbed for gravitational research.

Free space, i.e., gravitationally unaffected vacuum,   is the reference medium in electromagnetics \cite{Post}. A  metric with $g_{0\ell}=0\,\forall \ell\in\lec1,2,3\ric$  identifies the 0th coordinate as time and delinks it from the remaining three coordinates (space); hence, the equivalent bianisotropic continuum is  an anisotropic dielectric--magnetic which is devoid of magnetoelectric properties (i.e., $\#\Gamma=\#0$) and is impedance-matched to free space. What would happen if a metric were such that its equivalent bianisotropic continuum is  free space endowed with magnetoelectric gyrotropy (i.e., $\#\Gamma\ne\#0$)? This communication arose from an attempt to answer that question.

Let $u$ and $w_{1,2,3}$ be four real scalars of which only $u$ is constrained to be  non-zero and positive. Let these four scalars be used to  construct the metric
\begin{equation}
\les{g_{\alpha\beta}}\ris= u^{-1}\left(u+w_1^2+w_2^2+w_3^2\right)^{-1/4}
\les
\begin{array}{cccc}
1 & w_1 & w_2 & w_3\\
w_1 & -u & 0 & 0\\
w_2 & 0 &-u & 0\\
w_3 & 0 & 0 & -u
\end{array}\ris\,.
\label{metric1}
\end{equation}
Then, $\bar{g}u^2=-1$ and   Eqs.~\r{conrel0} turn out to be
\begin{equation}
\label{conrel1}
\left.\begin{array}{l}
{\bf D} =\epso {\bf E} -\co^{-1}({\bf w}\times\=I)\.{\bf H}
\\
{\bf B} =\muo {\bf H} +\co^{-1}({\bf w}\times\=I)\.{\bf E}
\end{array}\right\}\,,
\end{equation}
where the magnetoelectric gyrotropy vector $\#w=w_1\ux+w_2\uy+w_3\uz$ in the Cartesian coordinate system. Clearly, the bianisotropic continuum equivalent to the metric \r{metric1} is like free space  with magnetoelectric properties.

If an object  made of a linear homogeneous material described by Eqs.~\r{conrel1} were to be placed in conventional free space, its scattering characteristics should depend on the magnitude and direction of $\#w$. We decided to theoretically examine this proposition by considering the scattering of light by a sphere made of this material. For that purpose, we employed a recently formulated analytic procedure that relies on closed-form vector spherical wavefunctions
 for an orthorhombic dielectric-magnetic material
with   magnetoelectric gyrotropy \cite{LM2012}. In this procedure, a transition matrix (commonly called the ``T matrix")
describes scattering by the homogeneous sphere made of the chosen material.

The derivation of the T matrix for general nonspherical scatterers being available \cite{LM2012}, we provide essential expressions and final results in Sec.~\ref{theory}. Section~\ref{results} presents numerical results to explicate the effects of magnetoelectric gyrotropy on the scattering of an incident plane wave. Special attention is paid to
 total scattering efficiency, the forward scattering efficiency, and the backscattering efficiency
as functions of (i) the size parameter of the sphere and (ii) the magnitude and direction of $\#w$
in relation to  the incident plane wave.
The
dependency $\exp(-i\omega t)$ on time $t$ is present but suppressed,  $\ko=\omega/\co$
is the free-space wave number,
and $\etao=\sqrt{\muo/\epso}$ is the intrinsic impedance of free space.  The asterisk denotes the complex conjugate.

\section{Theory}\label{theory}

Suppose that the  center of a homogeneous sphere of radius $a$ and made of a material with constitutive relations \r{conrel1} is located at the origin of a Cartesian coordinate system $(x,y,z)$. The ambient medium is free space. The sphere is illuminated by a plane wave with field phasors
\begin{equation}
\label{incEH}
\left.\begin{array}{l}
{\bf E}\inc({\bf r}) = \#e\inc\exp{(i  \#k\inc\.\#r)}\\
{\bf H}\inc({\bf r}) = \#h\inc  \exp{(i\#k\inc\.\#r)}
\end{array}\right\}\,.
\end{equation}
Without any loss of generality, we fix $\#k\inc=\ko\uz$, $\#e\inc\parallel\ux$, and $\#h\inc=\left(\omega\muo\right)^{-1}\left(\#k\inc\times\#e\inc\right)\parallel \uy$.

\subsection{Incident-field representation}
In order to formulate the T matrix, we must represent the incident field phasors \r{incEH} in terms of the  vector spherical wavefunctions defined for free space as
\cite{Stratton1941}
\begin{eqnarray}
&&
\left.\begin{array}{l}
{{\bf M}_{emn}^{(1)}}(\ko{\bf r}) = \nabla\times\left[\#r j_n(\ko{r})P_n^m(\cos\theta)\cos(m\phi)\right]
\\[5pt]
{{\bf M}_{omn}^{(1)}}(\ko{\bf r}) = \nabla\times\left[\#r j_n(\ko{r})P_n^m(\cos\theta)\sin(m\phi)\right]
\\[5pt]
{{\bf N}_{smn}^{(1)}}(\ko{\bf r})=\ko^{-1}\nabla\times{{\bf M}_{smn}^{(1)}}(\ko{\bf r})\,,\quad s\in\left\{e,o\right\}
\end{array}
\right\}\,,
\nonumber
\\[5pt]
&&\qquad\qquad \qquad\qquad
m\in\lec0,1,2,...,n\ric\,,\quad n\in\lec1,2,3,...\ric\,,
\end{eqnarray}
with $j_n(\.)$ denoting the spherical Bessel function of order $n$, and $P_n^m(\.)$ the associated
Legendre function of order $n$ and degree $m$. The spherical coordinate system $(r,\theta,\phi)$ is
equivalent to the Cartesian coordinate system $(x,y,z)$. The  expansions \cite{Stratton1941,BH1983}
\begin{eqnarray}
&&\nonumber
{\bf E}\inc({\bf r}) = \sum^{\infty}_{n=1}
\left\{i^n{\frac {2n+1} {n(n+1)}}
\left[
{{\bf M}_{o1n}^{(1)}}(\ko{\bf r})\right.\right.
\\
&&\qquad\qquad\qquad\left.\left.-i
{{\bf N}_{e1n}^{(1)}}(\ko{\bf r})\right]\right\}\,, \label{incE1}
\\
&&\nonumber
{\bf H}\inc({\bf r}) = \frac{1}{i\etao}\,  \sum^{\infty}_{n=1}
\left\{i^n{\frac {2n+1} {n(n+1)}}
\left[
{{\bf N}_{o1n}^{(1)}}(\ko{\bf r})\right.\right.\\
&&\qquad\qquad\qquad\left.\left.-i{{\bf M}_{e1n}^{(1)}}(\ko{\bf r})\right]\right\}\,,
\label{incH1}
 \end{eqnarray}
follow from Eqs.~\r{incEH}.

However, as the scattering sphere is made of a bianisotropic material, it is  convenient to recast Eqs.~\r{incE1}
and \r{incH1} more generally as \cite{LM2012}
\begin{eqnarray}
&&\nonumber
{\bf E}\inc({\bf r}) = \sum_{s\in\left\{e,o\right\}}\sum^{\infty}_{n=1}\sum^n_{m=0}
\left\{D_{mn}\left[A_{smn}^{(1)}\,
{{\bf M}_{smn}^{(1)}}(\ko{\bf r})
\right.\right.\\
&&\qquad\qquad\qquad\left.\left.+
B_{smn}^{(1)}\,
{{\bf N}_{smn}^{(1)}}(\ko{\bf r})\right]\right\}\,,
\label{incE2}
\\&&
\nonumber
{\bf H}\inc({\bf r}) = \frac{1}{i\etao}\,\sum_{s\in\left\{e,o\right\}}\sum^{\infty}_{n=1}\sum^n_{m=0}
\left\{D_{mn}\left[A_{smn}^{(1)}\,
{{\bf N}_{smn}^{(1)}}(\ko{\bf r})
\right.\right.\\
&&\qquad\qquad\qquad\left.\left.+
B_{smn}^{(1)}\,
{{\bf M}_{smn}^{(1)}}(\ko{\bf r})\right]\right\}\,,
\label{incH2}
\end{eqnarray}
where the normalization factor
\begin{equation}
D_{mn}=(2-\delta_{m0}){(2n+1)(n-m)!\over 4n(n+1)(n+m)!}
\end{equation}
employs the Kronecker delta $\delta_{mn}$, and the coefficients
\begin{equation}
\left.\begin{array}{l}
{ A_{smn}^{(1)} = i^n {\frac {2n+1} {D_{mn}n(n+1)}}\delta_{m1}\delta_{so}},\\[5pt]
{B_{smn}^{(1)} = -i^{n+1} {\frac {2n+1} {D_{mn}n(n+1)}}\delta_{m1}\delta_{se}}
\end{array}\right\}\,.
\end{equation}

\subsection{Scattered-field representation}
The scattered electric and magnetic field phasors are represented as \cite{LM2012}
\begin{eqnarray}
&&\nonumber
{\bf E}\sca({\bf r}) = \sum_{s\in\left\{e,o\right\}}\sum^{\infty}_{n=1}\sum^n_{m=0}
\left\{D_{mn}\left[A_{smn}^{(3)}\,
{{\bf M}_{smn}^{(3)}}(\ko{\bf r})
\right.\right.\\
&&\qquad\qquad\qquad\left.\left.+
B_{smn}^{(3)}\,
{{\bf N}_{smn}^{(3)}}(\ko{\bf r})\right]\right\}\,, \quad r\geq a\,,
\label{Esca-def}
\\&&
\nonumber
{\bf H}\sca({\bf r}) = \frac{1}{i\etao}\,\sum_{s\in\left\{e,o\right\}}\sum^{\infty}_{n=1}\sum^n_{m=0}
\left\{D_{mn}\left[A_{smn}^{(1)}\,
{{\bf N}_{smn}^{(3)}}(\ko{\bf r})
\right.\right.\\
&&\qquad\qquad\qquad\left.\left.+
B_{smn}^{(3)}\,
{{\bf M}_{smn}^{(3)}}(\ko{\bf r})\right]\right\}\,, \quad r\geq a\,.
\label{Hsca-def}
\end{eqnarray}
In these expressions, the vectors spherical wavefunctions
${{\bf M}_{smn}^{(3)}}(\ko{\bf r})$ and ${{\bf M}_{smn}^{(3)}}(\ko{\bf r})$, respectively,
are defined the same way as ${{\bf M}_{smn}^{(1)}}(\ko{\bf r})$ and ${{\bf M}_{smn}^{(1)}}(\ko{\bf r})$,
except that the spherical Bessel function $j_n(\.)$ is replaced by
  the spherical Hankel function  $h^{(1)}_n(\.)$ of the first kind \cite{Stratton1941}.
The coefficients
$A_{smn}^{(3)}$ and $B_{smn}^{(3)}$ have to be determined by the solution of a boundary-value problem \cite{LM2012}.

In the far zone,  the scattered electric field  may be approximated as
\begin{equation}
\#E\sca({\#r})\approx\#F\sca(\theta,\phi) \frac{\exp(i{\ko}r)}{r}\,
\end{equation}
and the scattered magnetic field as
\begin{equation}
\#H\sca({\#r})\approx\etao^{-1}\ur\times\#F\sca(\theta,\phi) \frac{\exp(i{\ko}r)}{r}\,,
\end{equation}
where $\hat{\#r}=\#r/r$ and $\#F\sca(\theta,\phi)$ is the vector far-field scattering amplitude \cite{JLjosa1}.
The   differential scattering efficiency is given by
\begin{equation}\label{sigmad-def}
Q_{\rm D}(\theta,\phi)=\frac{4}{a^2}\,\frac{\#F\sca(\theta,\phi)\.\#F\sca^\ast(\theta,\phi)}{\#e\inc\.\#e\inc^\ast}\,.
\end{equation}
and total scattering efficiency is given as
\begin{eqnarray}
\label{sigmasca-def}
\sigma\sca&=&\frac{1}{\#e\inc\.\#e\inc^\ast}\,\int_{\phi=0}^{2\pi}\int_{\theta=0}^{\pi}\,
\left[{\#F\sca(\theta,\phi)\.\#F\sca^\ast(\theta,\phi)}\right]\sin\theta\,d\theta\,d\phi\,
\\
Q\sca&=&
\label{Qsca}
\frac{1}{\#e\inc\.\#e\inc^\ast}\,\frac{1}{{(\ko a)}^2}\,
\sum_{s\in\left\{e,o\right\}}\sum^{\infty}_{n=1}\sum^n_{m=0}\left[
D_{mn}\left(\vert A\smn^{(3)}\vert^2+ \vert B\smn^{(3)}\vert^2\right)\right]\,.
\end{eqnarray}

\subsection{Internal-field representation}
The electric and magnetic field phasors excited inside the vacuum-like sphere with
magnetoelectric gyrotropy are represented by \cite{LM2012,LWmotl1997}
\begin{eqnarray}
&&\nonumber
{\bf E}\exc({\bf r}) = \exp{(i\ko{\#w}\.{\# r})} \sum_{s\in\left\{e,o\right\}}\sum^{\infty}_{n=1}\sum^n_{m=0}
 \left[b_{smn}\,
{{\bf M}_{smn}^{(1)}}(\ko{\bf r})
\right.\\
&&\qquad\qquad\qquad\left.+
c_{smn}\,
{{\bf N}_{smn}^{(1)}}(\ko{\bf r})\right]\,, \quad r\leq a\,,
\label{Eint}
\\&&
\nonumber
{\bf H}\exc({\bf r}) = \frac{1}{i\etao}\,\exp{(i\ko{\#w}\.{\# r})} \sum_{s\in\left\{e,o\right\}}\sum^{\infty}_{n=1}\sum^n_{m=0}
 \left[b_{smn}\,
{{\bf N}_{smn}^{(1)}}(\ko{\bf r})
\right.\\
&&\qquad\qquad\qquad\left.+
c_{smn}\,
{{\bf M}_{smn}^{(1)}}(\ko{\bf r})\right]\,, \quad r \leq a\,,
\label{Hint}
\end{eqnarray}
 the coefficients $b_{smn}$ and $c_{smn}$ being unknown.

\subsection{Solution of boundary-value problem}
The standard boundary conditions
\begin{equation}
\left.\begin{array}{l}
\ur\times\#E\exc(\#r)=\ur\times\les\#E\inc(\#r)+\#E\sca(\#r)\ris
\\
\ur\times\#H\exc(\#r)=\ur\times\les\#H\inc(\#r)+\#H\sca(\#r)\ris
\end{array}\right\}\,,\quad
{r=a}\,,
\end{equation}
hold across the surface of the sphere.
Their application yields the following set of algebraic equations
for every combination of $j\in\lec1,3\ric$, $s\in\lec{e,o}\ric$, $n\in\lec1,2,3,...\ric$, and $m\in\lec0,1,2,...,n\ric$\cite{LM2012}:
\begin{eqnarray}
\label{Asmn1}
&&A_{smn}^{(1)}= \sum_{{s^\prime}\in\left\{e,o\right\}}\sum^{\infty}_{{n^\prime}=1}\sum^{n^\prime}_{{m^\prime}=0}\les
I_{smn,{s^\prime}{m^\prime}{n^\prime}}^{(1)}\,b_{{s^\prime}{m^\prime}{n^\prime}}+
J_{smn,{s^\prime}{m^\prime}{n^\prime}}^{(1)}\,c_{{s^\prime}{m^\prime}{n^\prime}}
\ris\, ,
\\[8pt]
&&
\label{Bsmn1}
B_{smn}^{(1)}= \sum_{{s^\prime}\in\left\{e,o\right\}}\sum^{\infty}_{{n^\prime}=1}\sum^{n^\prime}_{{m^\prime}=0}\les
{J}_{smn,{s^\prime}{m^\prime}{n^\prime}}^{(1)}\,b_{{s^\prime}{m^\prime}{n^\prime}}+
{I}_{smn,{s^\prime}{m^\prime}{n^\prime}}^{(1)}\,c_{{s^\prime}{m^\prime}{n^\prime}}
\ris\,,
\\[8pt]
\label{Asmn3}
&&A_{smn}^{(3)}= -\sum_{{s^\prime}\in\left\{e,o\right\}}\sum^{\infty}_{{n^\prime}=1}\sum^{n^\prime}_{{m^\prime}=0}\les
I_{smn,{s^\prime}{m^\prime}{n^\prime}}^{(3)}\,b_{{s^\prime}{m^\prime}{n^\prime}}+
J_{smn,{s^\prime}{m^\prime}{n^\prime}}^{(3)}\,c_{{s^\prime}{m^\prime}{n^\prime}}
\ris\, ,
\\[8pt]
\label{Bsmn3}
&&
B_{smn}^{(3)}= -\sum_{{s^\prime}\in\left\{e,o\right\}}\sum^{\infty}_{{n^\prime}=1}\sum^{n^\prime}_{{m^\prime}=0}\les
{J}_{smn,{s^\prime}{m^\prime}{n^\prime}}^{(3)}\,b_{{s^\prime}{m^\prime}{n^\prime}}+
{I}_{smn,{s^\prime}{m^\prime}{n^\prime}}^{(3)}\,c_{{s^\prime}{m^\prime}{n^\prime}}
\ris\,.
\end{eqnarray}
{In these equations,
the quantities
$I_{smn,{s^\prime}{m^\prime}{n^\prime}}^{(j)}$ and
$J_{smn,{s^\prime}{m^\prime}{n^\prime}}^{(j)}$
are computed as surface integrals. Thus,}
\begin{eqnarray}
\nonumber
I_{smn,{s^\prime}{m^\prime}{n^\prime}}^{(j)}=&& \frac{i(\ko a)^2}{\pi}
\int_0^{2\pi} d\phi\int_0^{\pi} d\theta\,\sin\theta
\lec\ur\.
\les
\#M\smn^{(\ell)}(\ko a\ur)
\times   \#N^{(1)}_{{s^\prime}{m^\prime}{n^\prime}}({\ko}a\ur)
\right.\right.
\\
&&\left.\left.
+\#N\smn^{(\ell)}(\ko a\ur)
\times
\#M^{(1)}_{{s^\prime}{m^\prime}{n^\prime}}({\ko}a\ur)
\ris
\exp(i{\ko}a{\#w}\.\ur)\ric\,
\label{Ismn}
\end{eqnarray}
and
\begin{eqnarray}
\nonumber
J_{smn,{s^\prime}{m^\prime}{n^\prime}}^{(j)}=&& \frac{i(\ko a)^2}{\pi}
\int_0^{2\pi} d\phi\int_0^{\pi} d\theta\,\sin\theta
\lec\ur\.
\les
\#M\smn^{(\ell)}(\ko a\ur)
\times   \#M^{(1)}_{{s^\prime}{m^\prime}{n^\prime}}({\ko}a\ur)
\right.\right.
\\
&&\left.\left.
+\#N\smn^{(\ell)}(\ko a\ur)
\times
\#N^{(1)}_{{s^\prime}{m^\prime}{n^\prime}}({\ko}a\ur)
\ris
\exp(i{\ko}a{\#w}\.\ur)\ric\,,
\label{Ismn}
\end{eqnarray}
where $j\in\lec1,3\ric$, $\ell = j+2\, (\mbox{mod} \,4)\in\lec3,1\ric$, and $\ur=\left(\ux\cp+\uy\sp\right)\st+\uz\ct$ is the unit radial vector. 
The integrals over $\phi$ can be handled analytically, but it is more convenient to evaluate
them numerically. The integrals over $\theta$ require numerical integration.
Let us also note that
$I_{smn,{s^\prime}{m^\prime}{n^\prime}}^{(3)}=-I_{{s^\prime}{m^\prime}{n^\prime},smn}^{(3)}$,
$J_{smn,{s^\prime}{m^\prime}{n^\prime}}^{(3)}=-J_{{s^\prime}{m^\prime}{n^\prime},smn}^{(3)}$,
and
$I_{smn,smn}^{(3)}=J_{smn,smn}^{(3)}=0$,
but similar skew-symmetric features are not displayed, in general, by
$I_{smn,{s^\prime}{m^\prime}{n^\prime}}^{(1)}$ and
$J_{smn,{s^\prime}{m^\prime}{n^\prime}}^{(1)}$.

The summations over $n^\prime\in\lec1,2,3,...\ric$  are restricted to $n^\prime\in\lec1,2,3,...,N\ric$ and similarly $n\in\lec1,2,3,...\ric$ to $n\in\lec1,2,3,...,N\ric$. Then,
Eqs.~\r{Asmn1}--\r{Bsmn3} can be written down symbolically in matrix form as
{\begin{equation}
\label{Y1def}
\les\begin{array}{c}
\les{A^{(1)}}\ris\\\les{B^{(1)}}\ris
\end{array}\ris
=
\les\begin{array}{cc}
\les{I^{(1)}}\ris  & \les{J^{(1)}}\ris \\
\les{J^{(1)}}\ris  & \les{I^{(1)}}\ris
\end{array}\ris
\les\begin{array}{c}
\les{b}\ris\\\les{c}\ris
\end{array}\ris
\equiv
\les{Y^{(1)}}\ris
\les\begin{array}{c}
\les{b}\ris\\\les{c}\ris
\end{array}\ris
\end{equation}
and
\begin{equation}
\label{Y3def}
\les\begin{array}{c}
\les{A^{(3)}}\ris\\\les{B^{(3)}}\ris
\end{array}\ris
=-
\les\begin{array}{cc}
\les{I^{(3)}}\ris  & \les{J^{(3)}}\ris \\
\les{J^{(3)}}\ris  & \les{I^{(3)}}\ris
\end{array}\ris
\les\begin{array}{c}
\les{b}\ris\\\les{c}\ris
\end{array}\ris
\equiv
-\les{Y^{(3)}}\ris
\les\begin{array}{c}
\les{b}\ris\\\les{c}\ris
\end{array}\ris
\,.
\end{equation}
Here, the column vectors
$\les{A^{(j)}}\ris$ and $\les{B^{(j)}}\ris$   contain  the coefficients
$A_{smn}^{(j)}$ and $B_{smn}^{(j)}$, respectively, arranged
in a specified order, with similar interpretations for the column vectors $\les{b}\ris$   and
$\les{c}\ris$. Furthermore, $\les{I^{(j)}}\ris$ and $\les{J^{(j)}}\ris$ are matrixes in which
the integrals $I_{smn,{s^\prime}{m^\prime}{n^\prime}}^{(j)}$ and
$J_{smn,{s^\prime}{m^\prime}{n^\prime}}^{(j)}$, respectively, are arranged in consonance
with the column vectors $\les{A^{(j)}}\ris$, etc.}

Equations~\r{Y1def} and \r{Y3def} lead to the relation
\begin{equation}
\les\begin{array}{c}\les{A^{(3)}}\ris\\
\les{B^{(3)}}\ris\end{array}\ris
=\les{T}\ris\,
\les\begin{array}{c}\les{A^{(1)}}\ris\\
\les{B^{(1)}}\ris\end{array}\ris
\,,
\end{equation}
wherein $\les{T}\ris=-\les{Y^{(3)}}\ris\,\les{Y^{(1)}}\ris^{-1}$ is the T matrix of the chosen sphere suspended in free space. Because of the structure of the matrix $\les{Y^{(j)}}\ris$, the T matrix
can be partitioned as
\begin{equation}
\les{T}\ris\equiv
\les\begin{array}{cc}
\les{T^{(A)}}\ris  & \les{T^{(AB)}}\ris \\
\les{T^{(AB)}}\ris  & \les{T^{(B)}}\ris
\end{array}\ris\,.
\end{equation}

\section{Numerical results and discussion} \label{results}
We set up a Mathematica\texttrademark~program to compute the T matrix. In the program, we truncated the summations over $n^\prime\in\lec1,2,3,...\ric$ to $n^\prime\in\lec1,2,3,...,N\ric$ and similarly $n\in\lec1,2,3,...\ric$ to $n\in\lec1,2,3,...,N\ric$. We chose sufficiently high values of $N$,
such that the extinction efficiency $Q\ext$, the total scattering efficiency $Q\sca$, the forward scattering efficiency $Q_f$, and the backscattering efficiency $Q_b$ \cite{JLjosa1} converged to a pre-set tolerance of
0.1\%. Smaller values of $\vert\#w\vert$ and ${\ko}a$
required smaller $N$, with $N=11$ being adequate for $\vert\#w\vert=0.25$ and $\ko a=4.0$.

We   confirmed that our program yielded negligibly tiny values of
the coefficients $A_{smn}^{(3)}$ and $B_{smn}^{(3)}$
when we set $\#w=\#0$. When $\#w\perp\uz$, reversal of the direction of $\#w$ was tantamount to the multiplication of $A_{smn}^{(3)}$, $B_{smn}^{(3)}$,
$b_{smn}$, and $c_{smn}$ by negative unity, which left $Q\ext$, $Q\sca$, $Q_f$, and $Q_b$ unchanged. When $\#w\parallel\uz$,  both $\#w$ and the direction of propagation of the incident plane wave had to be reversed together for  $Q\ext$, $Q\sca$, $Q_f$, and $Q_b$ to remain unchanged.
Regardless of the choice of $\#w$, we found that $Q\ext=Q\sca$, implying that the absorption efficiency $Q\abs=0$.   This was expected because Eqs.~\r{conrel1} satisfy the conditions of the absence of dissipation \cite{Tellegen,MLbook}.

Although the magnetoelectric gyrotropy vector $\#w$ can be arbitrarily oriented, three cases are of particular interest because the incident light is a plane wave:
\begin{itemize}
\item $\#w$ is parallel to the incident electric field (i.e., $\#w\parallel\ux$),
\item $\#w$ is parallel to the incident magnetic field (i.e., $\#w\parallel\uy$), and
\item $\#w$ is parallel to the propagation vector of the incident plane wave (i.e., $\#w\parallel\uz$).
\end{itemize}

\subsection{Efficiencies}
\begin{figure}[!htbp]
\centering
{\includegraphics[width=6.6in]{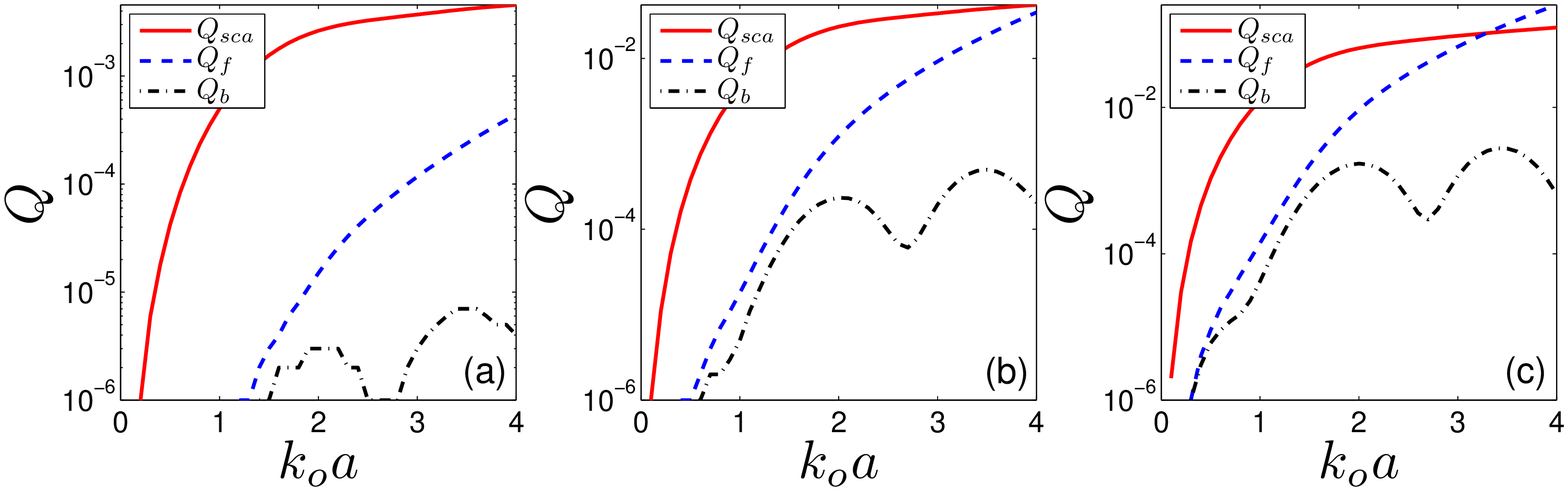}}
\caption{$Q\sca$, $Q_f$, and $Q_b$ as functions of the size parameter ${\ko}a$, when $\#w$ is parallel
to the incident electric field;
$w_2=w_3=0$, but
(a) $w_1=0.05$, (b) $w_1=0.15$, and (c) $w_1=0.25$. These plots also hold true
when $\#w$ is parallel
to the incident magnetic field; $w_1=w_3=0$, but
(a) $w_2=0.05$, (b) $w_2=0.15$, and (c) $w_2=0.25$.
}
\label{fig1}
\end{figure}

Figure~\ref{fig1} shows plots of $Q\sca$, $Q_f$, and $Q_b$ as functions of the size parameter ${\ko}a$ when
$\#w\parallel\#e\inc$ and $\vert\#w\vert \in\lec0.05,0.15,0.25\ric$. An
increase in the magnitude of the magnetoelectric gyrotropy vector has a more pronounced effect on $Q_f$ than on $Q\sca$ and $Q_b$. Whereas
$Q_{sca}$ is higher than $Q_{f}$ for smaller values of $\vert\#w\vert$ and ${\ko}a$,   the reverse is true for larger values of $\vert\#w\vert$ and ${\ko}a$. The backscattering efficiency shows oscillatory behavior and peaks of the oscillations increase as the  size parameter  increases.

The plots of $Q\sca$, $Q_f$, and $Q_b$ as functions of the size parameter ${\ko}a$ when $\#w\parallel\#h\inc$  are identical to those when $\#w\parallel\#e\inc$. Thus, the effect of magnetoelectric gyrotropy is independent of its orientation when $\#w\perp \#k\inc$.

\begin{figure}[!htbp]
\centering
{\includegraphics[width=6.6in]{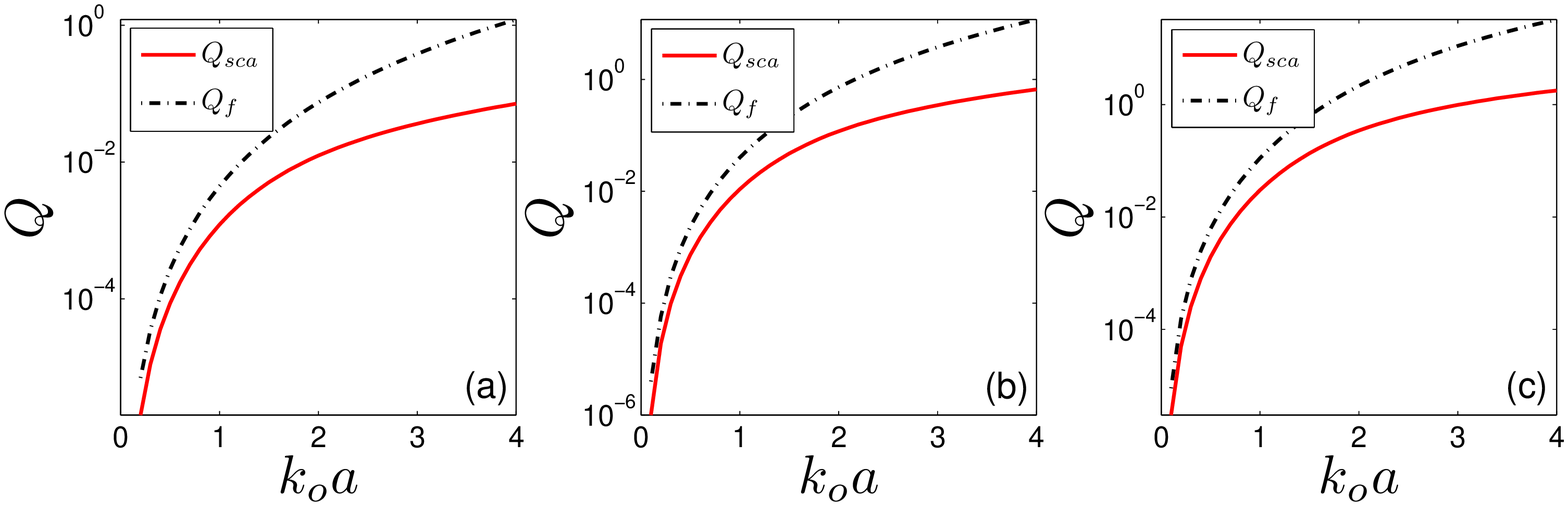}}
\caption{$Q\sca$ and $Q_f$  as functions of the size parameter ${\ko}a$, when $\#w$ is parallel
to the direction of propagation of the incident plane wave;
$w_1=w_2=0$, but (a) $w_3=0.05$, (b) $w_3=0.15$, and (c) $w_3=0.25$. $Q_b\equiv0$ when $\#w\parallel\#k\inc$.}
\label{fig2}
\end{figure}

Figure~\ref{fig2} shows plots of $Q\sca$, $Q_f$, and $Q_b$ as functions of the size parameter ${\ko}a$ when
$\#w\parallel\#k\inc$ and $\vert\#w\vert \in\lec0.05,0.15,0.25\ric$.
The  influence of   magnetoelectric gyrotropy  is maximal when $\#w\parallel\#k\inc$,
as is evident from a comparison of Figs.~\ref{fig1}
and \ref{fig2}.
The maximum value of $Q\sca$ is an order of magnitude higher and
$Q_f$ in Fig.~\ref{fig1} is two orders of magnitude higher when
$\#w\parallel\#k\inc$
 than
when ${\#w} \perp {\#k\inc}$. Moreover, there is no backscattering   (i.e., $Q_b=0$) when
$\#w\parallel\#k\inc$, which makes the sphere invisible in the monostatic configuration.

The absence of backscattering when $\#w\parallel\#k\inc$ has an analog \cite{BohrenAO1988}
in the reflection of a plane wave incident normally at the planar interface
of free space and the material with constitutive relations \r{conrel1} such that
$\#w$ is oriented wholly normal to the interface. Simple algebraic manipulations show
that reflection is then absent (and transmission is perfect).

\begin{figure}[!htbp]
\centering
{\includegraphics[width=6.6in]{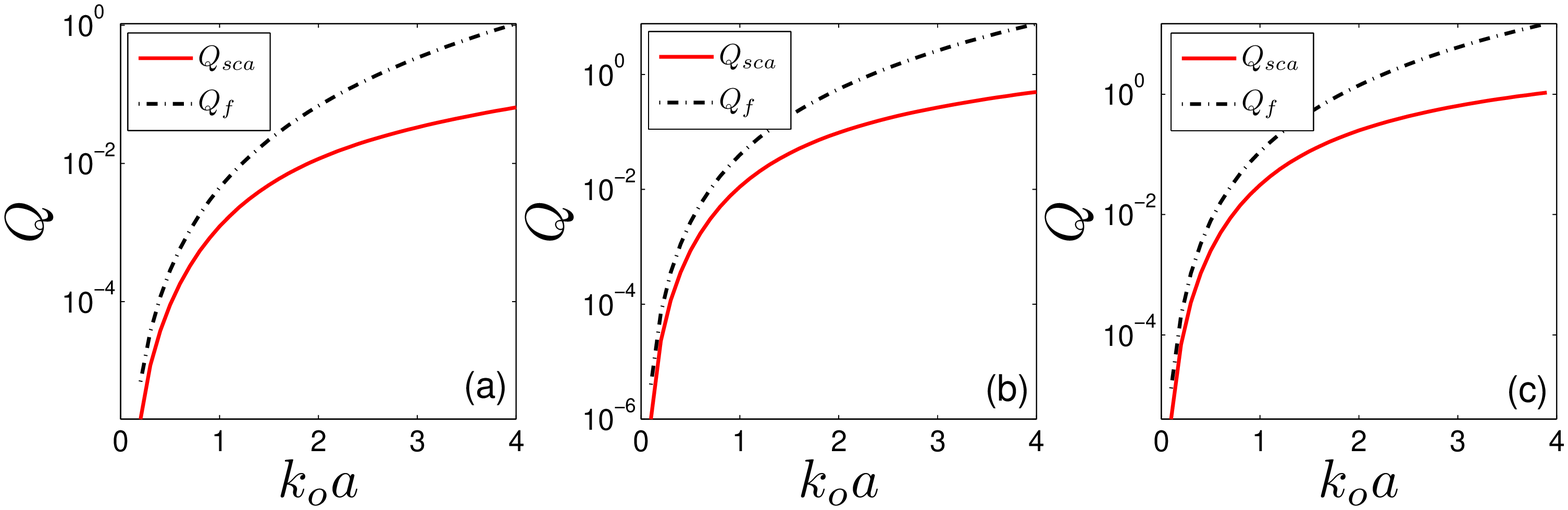}}
\caption{Same as Fig.~\ref{fig2}, except that (a) $w_3=-0.05$, (b) $w_3=-0.15$, and (c) $w_3=-0.25$.
}
\label{fig3}
\end{figure}

Figure~\ref{fig3} shows the same plots as Fig.~\ref{fig2}, except that $w_3<0$.
A change in the sign of $w_3$ affects both $Q\sca$ and $Q_f$, as is clear from comparing
 Figs.~\ref{fig2} and \ref{fig3}. Both
$Q\sca$ and $Q_f$ are higher when $\#w$ is coparallel, than when
$\#w$ is antiparallel,  to   the propagation vector $\#k\inc$ of the incident plane wave.

Given the foregoing trends, for arbitrarily directed $\#w$ it is reasonable to expect that the effects of the component of $\#w$ that is co/anti-parallel to the propagation vector of the incident plane wave would dominate those of the component of $\#w$ that is perpendicular to the propagation vector. Several calculations (not shown) validated that expectation.

\subsection{Differential scattering efficiency}
\begin{figure}[!htbp]
\centering
{\includegraphics[width=6.6in]{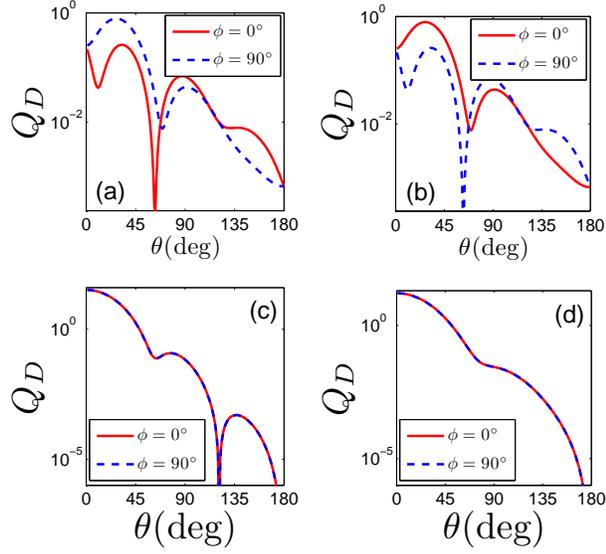}}
\caption{$Q_D$  as a function of {$\theta\in[0^\circ,180^\circ]$
and $\phi=\lec0^\circ,90^\circ\ric$} for $\ko a=4 $, when (a) $w_1=0.25$, $w_2=w_3=0$; (b) $w_1=w_3=0, w_2=0.25$;  (c) $w_1=w_2=0, w_3= 0.25$; {and (d) $w_1=w_2=0, w_3= -0.25$.}
}
\label{diffscfig}
\end{figure}

For $\ko{a}=4$, $\vert\#w\vert=0.25$, and four different orientations
of $\#w$, the differential scattering efficiencies $Q_D(\theta,0^\circ)$
and $Q_D(\theta,90^\circ)$ are plotted versus the observation angle
$\theta\in[0^\circ,180^\circ]$  in Fig.~$\ref{diffscfig}$. The curve of $Q_D(\theta,0^\circ)$  when $\#w\parallel\#e\inc$ [Fig. 4(a)] is identical
to that of $Q_D(\theta,90^\circ)$   when $\#w\parallel\#h\inc$ [Fig. 4(b)]. Likewise,
the curve of $Q_D(\theta,90^\circ)$  for $\#w\parallel\#e\inc$  is identical
to that of $Q_D(\theta,0^\circ)$   for $\#w\parallel\#h\inc$. This shows that the impact of magnetoelectric gyrotropy is largely independent of its orientation when $\#w\perp\#k\inc$. More lobes appear in
the curve of $Q_D(\theta,0^\circ)$ as compared to $Q_D(\theta,90^\circ)$, and
the maximum magnitude of the former is smaller than that of the latter, when  $\#w\parallel\#e\inc$.

When $\#w$ is co/anti-parallel to $\#k\inc$, the differential scattering appears identical in 
the $\phi=0^\circ$  and  $\phi=90^\circ$ planes, as shown in Figs. $\ref{diffscfig}$(c) and $\ref{diffscfig}$(d). However, more lobes exist when $\#w$ is co-parallel than when it is anti-parallel to $\#k\inc$.

\subsection{Rayleigh scattering}\label{Rs}
A long-wavelength approximation yields closed-form analytical results for scattering by homogeneous and
electrically small objects \cite{Lrayleigh}. Accordingly, Rayleigh scattering by the chosen sphere
is equivalent to radiation jointly by an electric dipole moment
\begin{equation}
\#p\eqvt=-\left(\frac{4\pi{a^3}}{3}\right)\frac{\epso}{1-\left(\vert{\#w}\vert/3\right)^2}
\left(\#w\times\=I\right)\.
\left[\left(\hat{\#k}\inc\times\=I\right)-\frac{1}{3}\left(\#w\times\=I\right)\right]\.\#e\inc
\end{equation}
and a magnetic dipole moment
\begin{equation}
\#m\eqvt=
\left(\frac{4\pi{a^3}}{3}\right)\frac{\sqrt{\muo\epso}}{1-\left(\vert{\#w}\vert/3\right)^2}
\left(\#w\times\=I\right)\.
\left[\=I+\frac{1}{3}\left(\#w\times\=I\right)\.\left(\hat{\#k}\inc\times\=I\right)\right]\.\#e\inc
\end{equation}
both located at the centroid of the sphere, with $\hat{\#k}\inc=\#k\inc/\ko$. Clearly from these
expressions, both equivalent dipole moments vanish as $\vert{\#w}\vert\to0$.

Therefore, the Rayleigh estimate of the vector far-field scattering amplitude
is \cite{JLjosa1,Lrayleigh}
\begin{equation}
\#F\sca^{\rm Rayleigh}(\hat{\#r})= -\frac{\omega^2\muo}{4\pi}
\left[\hat{\#r}\times\left(\hat{\#r}\times\#p\eqvt\right)+\etao^{-1}\,\hat{\#r}\times\#m\eqvt\right]
\,,
\label{rayleigh}
\end{equation}
wherefrom the Rayleigh estimates of the various  efficiencies were obtained as follows:
\begin{eqnarray}
\nonumber
&&Q\sca^{\rm Rayleigh}=\frac{8(\ko{a})^4}{3\left(\#w\.\#w-9\right)^2}\\
\nonumber
&&\qquad\times
\left[
\left(w_1^2+w_2^2\right)^2
+3\left(w_1^2+w_2^2\right)\left(w_3-1\right)\left(w_3-3\right)\right.\\
&&\qquad\quad\left.
+2w_3^2\left(w_3-3\right)^2\right]\,,\\[6pt]
&&
Q_f^{\rm Rayleigh}=\frac{4(\ko{a})^4}{\left(\#w\.\#w-9\right)^2}
 \left[w_1^2+w_2^2+2w_3\left(w_3-3\right)\right]^2\,,\\[6pt]
 &&
 Q_b^{\rm Rayleigh}=\frac{4(\ko{a})^4}{\left(\#w\.\#w-9\right)^2}
 \left(w_1^2+w_2^2\right)^2\,.
 \label{eq37}
\end{eqnarray}

Neither $w_1$ nor $w_2$ occur by themselves in the foregoing expressions, but always
as $w_1^2+w_2^2$. Therefore,
when $\#w\perp\#k\inc$, the three efficiencies
\begin{eqnarray}
&&Q\sca^{\rm Rayleigh}=\frac{8(\ko{a})^4}
{3\left(\#w\.\#w-9\right)^2} {(\#w\.\#w)}\left({\#w\.\#w+9}\right)
\,,
\\[5pt]
&&Q_f^{\rm Rayleigh}=Q_b^{\rm Rayleigh}= \frac{4(\ko{a})^4}
{\left(\#w\.\#w-9\right)^2} {\left(\#w\.\#w\right)^2}\,,
\end{eqnarray}
contain the quadratic form $\#w\.\#w$ and are invariant with respect to the orientation
of the magnetoelectric gyrotropy vector.

In contrast when $\#w\parallel\#k\inc$, the  efficiencies
\begin{eqnarray}
&&Q\sca^{\rm Rayleigh}=\frac{1}{3}Q_f^{\rm Rayleigh}=
\frac{16(\ko{a})^4w^2}{3(w+3)^2}
\end{eqnarray}
do depend on the orientation of the magnetoelectric gyrotropy vector.
However, from Eq.~(\ref{eq37}) it follows that
$Q_b^{\rm Rayleigh}=0$ does not.

\begin{figure}[!htbp]
\centering
{\includegraphics[width=6.6in]{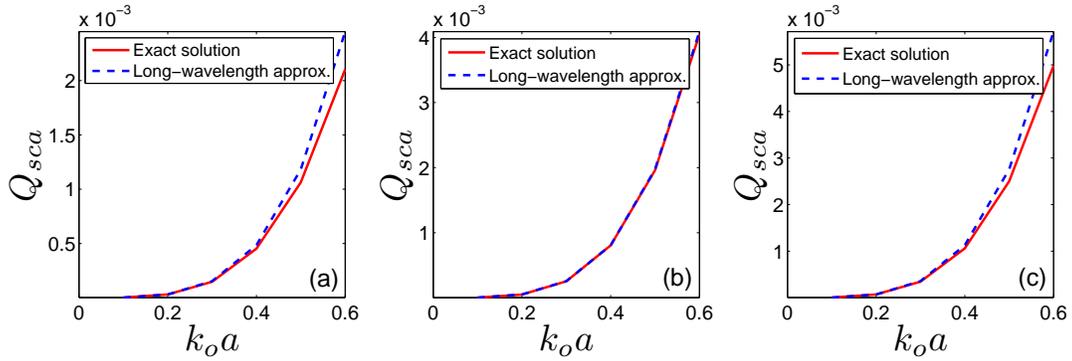}}
\caption{Total scattering efficiency computed exactly using Eq.~(\ref{Qsca}) and approximately using Eq.~(\ref{rayleigh}) as a function of $\ko a$ when (a) $ w_2=w_3=0, w_1=0.25$; (b) $ w_1=w_2=0, w_3=0.25$; and (c) $ w_1=w_2=0, w_3=-0.25$. }
\label{Rsfig}
\end{figure}

The Rayleigh  expressions are expected to hold when the radius of the sphere is less than a tenth of the free-space wavelength.
As an example, Fig.~\ref{Rsfig} depicts plots of
$Q\sca^{\rm Rayleigh}$ and $Q\sca$ versus $\ko{a}\in(0, 0.6]$ for $\vert{\#w}\vert=0.25$. Clearly, the long-wavelength approximation agrees well for entire range of $\ko a$ when $\#w$ is coparallel to $\#k\inc$. When $\#w$ is parallel to the incident electric/magnetic field or $\#w$ is antiparallel to $\#k\inc$, the results match well for $\ko{a}\in(0, 0.4]$ but the difference between the exact and approximate results begins to rise as the value of $\ko a $ increases beyond $0.4$.

\section{Concluding remarks}
Electromagnetic scattering by a vacuum-like sphere with   magnetoelectric gyrotropy was formulated in terms of the  T matrix,
after simplifying recently derived  vector spherical wavefunctions in closed form. The total scattering, extinction, forward scattering, and backscattering efficiencies were computed to explicate the magnitude and the direction of the magnetoelectric gyrotropy
vector in relation to the directions of the propagation vector, the magnetic field, and the electric field of a plane wave incident on the chosen sphere.

Since the  permittivity and the  permeability of the sphere are exactly the same as those of the surrounding vacuum, any scattering must be attributed solely to the magnetoelectric gyrotropy vector of the sphere. In general, all scattering efficiencies  grow as the magnetoelectric gyrotropy grows in magnitude. A growing trend in all  efficiencies with increase in the electrical size of the sphere was also found, though the growth may not be monotonic but undulatory.

Both the total scattering and forward scattering efficiencies are generally lower when the magnetoelectric gyrotropy vector of the sphere is perpendicular to the propagation vector of the incident plane wave than when it is anti-parallel to the propagation vector.
Further enhancements occur when  the magnetoelectric gyrotropy vector  is co-parallel to the propagation vector.   Furthermore,  the sphere is
invisible in monostatic configuration provided that the the magnetoelectric gyrotropy vector  is co/anti-parallel to the propagation vector.

\section*{Acknowledgments}
A.D.U.J.  thanks the Higher Education Commission of Pakistan  and  A.L. is grateful to the Charles Godfrey Binder Endowment at the Pennsylvania State University for  supporting this research.

\end{document}